# Atom Probe Tomography Spatial Reconstruction: Status and Directions

D. J. Larson[1,*], B. Gault[2], B. P. Geiser[1], F. De Geuser[3] and F. Vurpillot[4]

[1] CAMECA Instruments, Inc., 5500 Nobel Drive, Madison, WI 53711 USA

[2] Department of Materials, University of Oxford, Parks Road, Oxford, OX13PH UK

[3] SIMaP, Grenoble INP-UJF-CNRS, BP 75, 38402, Saint-Martin-d'Heres Cedex, France

[4] Groupe de Physique des Materiaux, UMR CNRS 6634, Université de Rouen, Saint Etienne du Rouvray 76801 France

* david.larson@ametek.com, tel : (608) 229-1938

## Abstract

In this review we present an overview of the current atom probe tomography spatial data reconstruction paradigm, and explore some of potential routes to improve the current methodology in order to yield a more accurate representation of nanoscale microstructure. Many of these potential improvement methods are directly tied to extensive application of advanced numerical methods, which are also very briefly reviewed. We have described effects resulting from the application of the standard model and then introduced several potential improvements, first in the far field, and, second, in the near field. The issues encountered in both cases are quite different but ultimately they combine to determine the spatial resolution of the technique.

## 1 INTRODUCTION

Over approximately the last decade, the performance and applicability of atom probe tomography (APT) has undergone a revolution [1] due to three factors: 1) improved specimen preparation due to focused ion beam milling, 2) improved field-of-view due to the advent of a micro- (i.e. local) electrode and other ion optical methods, and 3) reinvention of the use of lasers to induce field evaporation. This combination has created current challenges in the area of spatial data reconstruction algorithms for APT for two reasons. Firstly, datasets collected at wider field-of-view (e.g. 45°) are not accurately reconstructed using algorithms with small angle approximations. Secondly, laser pulsing enhances yield from specimens containing features such as interfaces and second phases, and these features are often difficult to reconstruct accurately.

In APT, ions are projected from the specimen onto a detector, and the properties of the projection are directly determined by the specimen itself. This is not the norm for most microscopies, and as the specimen is progressively analyzed, the field evaporating surface evolves in a complex manner [2, 3]. This produces specimen shapes which result in non-uniform magnification [4-7]. In the current work, variations which have the greatest effect very near the specimen surface are termed near field effects.

In the far field, beyond a distance corresponding to a few radii of curvature of the specimen, the ion trajectories are most influenced by the meso- and micro-scale parameters, such as the average radius curvature of the specimen, its shank angle, and the overall electrostatic environment of the specimen (i.e. presence of a substrate, distance to an extraction electrode, etc.) [8-13]. This intrinsic dichotomy between effects in the near and far field has long been known [14]. These effects in the near and far field combine to create challenges for the accurate reconstruction of APT data.

This paper is divided into four sections, the first of which summarizes the current reconstruction methodology, outlining the limitations and their causes. The second section minimally describes the computational methods employed extensively in the current work; for further information see [3, 7, 15-22]. The third and fourth sections present selected examples of our understanding of issues in the far and near field, respectively, and in each section, we have included the potential pathways to improve APT data reconstruction. The authors would like to note that we have recently published a review on the current topic [23] and have made no effort to reproduce those materials, but rather to build on that information and extend the presentation into the noted areas.

## 2 STATE-OF-THE-ART

### 2.1 *Single projection point, spherical shape and tangential continuity*

In atom probe tomography, the term *reconstruction* refers to the methodology enabling the transformation from detector ion-hit positions (X,Y,N) (where N is ion sequence number) into specimen spatial coordinates (x,y,z).

Most reconstruction algorithms have been designed based upon a set of assumptions similar to those included here [24-28]:

- the highly magnified image of the specimen surface formed by the successive impact of the ions on the detection is described by a point-projection;
- the field evaporated specimen is approximated by a spherical surface of radius R positioned on a truncated cone;

- the original location of the detected ion at the specimen surface may be computed by postulating the existence of a projection point *P*, located along the specimen main axis at a distance $\xi$R from the apex;
- either a magnification factor is estimated, allowing for conversion of (X,Y) to (x,y) positions [25], or the spatial coordinates (x,y,z) are defined as the intersection between the proposed ion emission surface and the straight line joining P and the detector hit position [27, 28];
- for each ion processed, the reconstructed specimen surface is moved by a small increment *dz* irrespective of the location of the impact on the detector (constant with respect to (X,Y)).

The main parameters used are *R*, the radius of curvature of the specimen, $\xi$, the image compression factor, and $\alpha$, the specimen shank angle. The depth increment *dz* is also a function of a number of instrumental and materials parameters such as flight-path length, detection efficiency, detector area, and atomic volume [25, 26].

It is generally accepted that the nominal value of the far field (e.g., not affected by crystallography or second phase materials contained within the specimen) image compression factor is between one and two. Its deviation from unity (away from a simple radial projection) is due to the effects of the shank region of the specimen compressing the electric field lines inwards toward the detector and thus falls into the category far field effects. The image compression factor is normally held constant over the entire field-of-view and throughout the whole analysis. The radius of curvature is determined either based on the evolution of the high voltage (*V)* applied to the specimen using a direct proportionality relationship between the field and the radius [29] ($F=V/k_f\,R$ where $k_f$ is normally between three and five) or, alternatively, constrained by the geometry of the specimen [30].

In the current work, a reconstruction method which is constrained by the assumptions presented up to this point will be termed the *standard model*. The main existing protocols may differ in the model assumed for the emitting surface, the estimation of the depth-increment, or the evolution of the radius of curvature [31]. Considering the hypotheses underpinning this model, it is expected to work well for limited field-of-view reconstructions, particularly for homogeneous materials, in which the single-point projection model accurately approximates the observed projection, as observed in field-ion-microscopy measurements [32, 33]. Well calibrated reconstructions based upon this method yield spatial resolution sufficiently high to image high Miller indices atomic planes in a variety of systems [34-39]; however, near-field effects still preclude imaging the full lattice, which may be partially retrieved using advanced data-treatment methods [36, 40-42].

## 2.2 *Position of the problem*

The standard model was originally developed for instruments with restricted fields-of-view, and like all models, it has some limitations. A simple, first-order, representation of the specimen surface as a spherical cap is insufficient for capturing the surface details that affect reconstruction accuracy in instruments that collect a large fraction of an emitter's surface area. It is known that with increasing field-of-view, the chances to image regions of the specimen with a surface curvature deviating from a spherical cap, which tends to yield distortions within the reconstruction [2, 3, 20, 22, 26, 27, 31, 43-53].

A field evaporated specimen shape may deviate from a hemispherical cap for a number of reasons: 1) An APT specimen is not a charged sphere in space, it is needle-shaped, and thus the effective electric field at the surface of the specimen is slightly higher at the apex than it is near the region where the evaporating surface intersects the shank of the needle. This effect results in a slight blunting near the apex center [20, 48, 49, 54-56]. In addition, experimental observations of specimens after field evaporation have long demonstrated that there is usually no tangential continuity between the conical shank and the emitting cap [44, 47, 49, 57-59], 2) the crystallographic dependence of evaporation field is strong in many materials/structures and will depend on the temperature at which the analysis is performed [15, 29, 44-46, 60-64], 3) different phases within a specimen may have significantly different evaporation fields [2, 3, 5, 7, 20-22, 52, 65-68], and 4) when laser pulses are used to trigger the field evaporation, temperature variations across a specimen surface may result in a surface curvature variation [69, 70].

These aspects are crucial due to the fundamental relationship between specimen radius of curvature, electric field, and magnification. Regions of low (high) radius of curvature produce regions of high (high) electric field and high (low) magnification locally [29, 71, 72]. The result is that, as the field-of-view increases, the single-point projection model increasingly fails to reproduce the non-constant magnification produced across the specimen surface (as demonstrated in the analysis of field-ion micrographs even in the simple case of pure metals [46, 73]). For the case of specimens of relative chemical inhomogeneity, this limitation of the standard model in the far field will be further developed as one of the main areas of potential development for evolving the reconstruction protocol to include a variable point-projection [20].

## 3 NUMERICAL METHODS OF INVESTIGATION

The complexity of the effects described above often makes it difficult to achieve analytical solutions to the specimen/instrument geometry found in atom probe tomography. Studies have hence relied on computational and numerical methods to investigate these problems in the near field (finite-element simulations) and in the far field (finite-element and boundary-element simulations). The methods used to generate the data presented herein are detailed in the following sections.

### 3.1 *Three-dimensional finite-element atomistic simulations*

The first method consists of a three-dimensional Poisson simulation using a finite difference algorithm and has been described in previously [19, 20] and is similar to other models [15, 21, 74]. For the study discussed herein, nested simulation volumes were used. The finest 3D grid had a resolution of 0.2 nm and extended for 100 nm laterally and 200 nm along the specimen axis. The coarsest 3D grid had a resolution of 0.8 nm and extended for 400 nm laterally and 800 nm along the specimen axis. Once propagating ions reached the upper edge of the simulation volume they were linearly projected along their current velocity direction to the detector plane, so the simulation may not accurately reproduce *global* system parameters such as far-field image compression which are also a function of instrument configuration. The specimen geometry was initially created with a spherical apex satisfying tangential continuity with the conical shank. In the current work, simulated data have a <001>-oriented face-centered-cubic single crystal structure with a lattice parameter of 0.404 nm (aluminum), an atomic density of 60 nm$^{-3}$, an initial specimen radius of 20 nm, and a shank angle of 5°.

### 3.2 *Two-dimensional finite-element atomistic simulations*

A simpler model recently developed by Vurpillot et al. [75] applies a dimensionality reduction by assuming cylindrical symmetry. The resultant simulations are less computationally expensive, which allows an expansion of the simulated space around the specimen to 10 cm x 20 cm, encompassing the microelectrode and the detector. The specimen consists of a compact stack of atomic planes, with an interatomic distance equal to a single unit cell (0.3 nm). By numerical calculation of the Poisson equation, the distribution of the electrostatic potential in the simulated space is computed on a mesh that is regular near the specimen and, beyond a limit (i.e. about 200 nm from the tip surface), the mesh size progressively increases with a geometric progression. In both model scenarios described above, the distribution of the potential is used to compute the electrostatic field, and ion trajectories may be derived.

### 3.3 *Two-dimensional boundary-element simulations*

As an alternative to the latter, the commercial particle trajectory analysis software LORENTZ 2D v9.0 was used to obtain the electric field distribution in a full-scale modern atom probe microscope (CAMECA LEAP® 3000X Si), slightly modified and reduced into an axially symmetric structure [11]. This software makes use of the boundary-element method, which is preferred over other volume-discretization methods for small surface-area-to-volume ratio and when a broad range of scales are covered. After the computation of the potential and electric field distributions, a fifth-order adaptive Runge-Kutta algorithm is used to compute the ion trajectories [11].

# 4 Results & Discussion: in the far field

## 4.1 *Limitations of a single-point projection*

It has long been known that the single point-projection provides a relatively poor description of field ion micrographs [46, 73] and atom probe data [76]. In both cases, data are better approximated by a projection where a linear relationship links the angle between two features at the specimen surface with the physical distance between their image on the screen or detector. The slope of this linear relationship is generally termed $k_\theta$. For many years this issue could justifiably be ignored (small field-of-view) but is more relevant for modern instruments with a large field-of-view.

At small angles, a direct relationship exists between the flight path (L) and the image compression factor (ξ): $k_\theta = L/\xi$. Hence by assuming a projection point it is possible to determine $k_\theta$. Assuming that the linear projection applies, we have analytically recomputed the positions of the points forming a plane within the original specimen of radius R=50nm in the case of a point-projection using both the approach of Bas et al. [25] and the more advanced approach of Geiser et al. [27]. The reverse projection of the plane in both cases is presented in Figure 1 (a). This approach was extended to a set of aligned spheres, and the resulting images through the reverse-projection using the different models are displayed in Figure 1 (b).

It is clear from this figure that the point-projection induces distortions due to its inability to reproduce the linear projection at distances far from x=0, which corresponds to the centre of the detector. The distortions induced onto the planes vary in amplitude and in sign depending on the projection used. More worrying is the distortion of the spheres that increase at higher angle with each sphere appearing elongated. This exercise demonstrates that even in cases where all the geometrical parameters describing the specimen are known, the standard reconstruction protocol does not yield a tomogram entirely free of distortions. The input parameters of the point-projection can be adjusted [27], as shown in light blue in Figure 1, so as to optimize the planarity of a flat feature, but even in this case there will be some distortion yielding unequal objects volumes along the x axis.

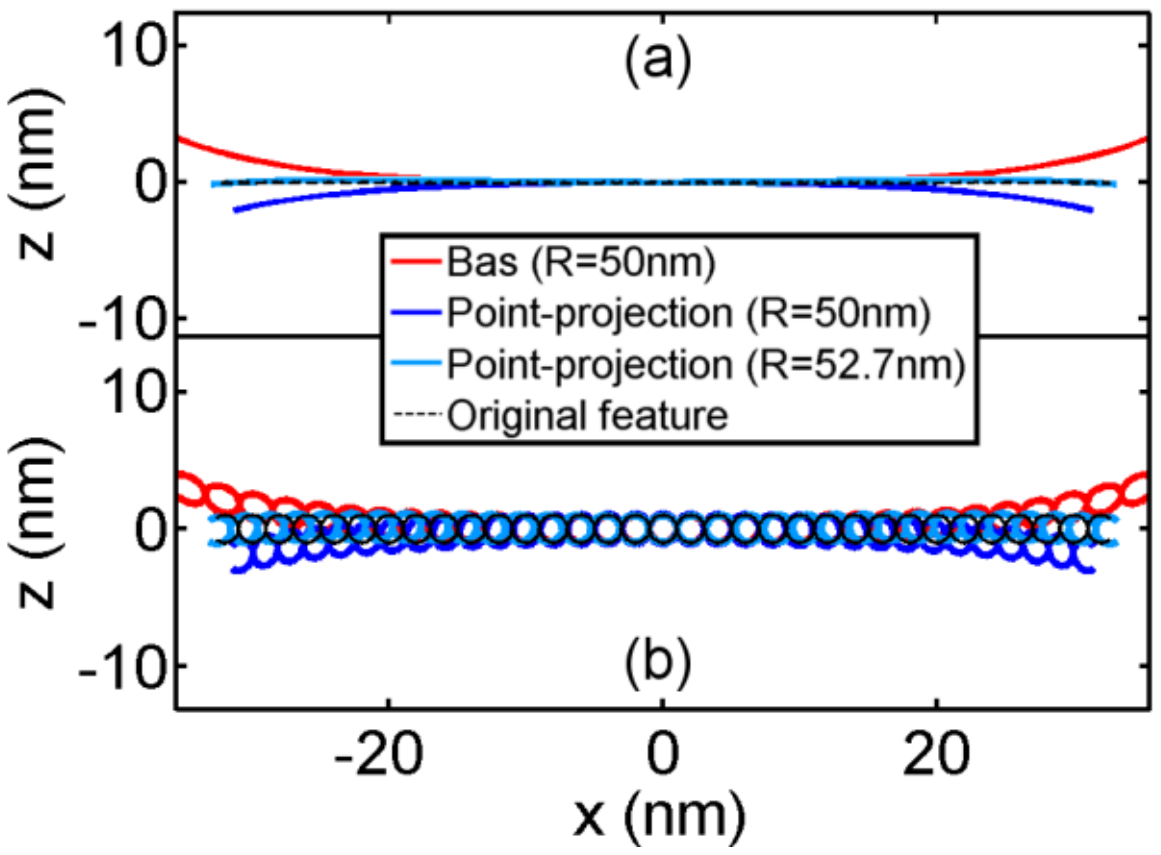

*Figure 1: (a) Reverse-projection of a plane (dashed line) in a specimen with a radius of 50 nm, (initially projected using a linear law with $k_\theta$=0.056 rad $m^{-1}$($\xi$=1.6, L=90 mm)), using the approximations of Bas et al. (red line), a true point projection with the input radius of curvature (dark blue line) and a point projection with a radius of curvature artificially increased to minimize the error (light blue line). Note the difference in scale between x and z. (b) Similar approach extended to a set of aligned spheres.*

Use of an inappropriate projection to create a reconstruction is, of course, expected to cause dimensional artifacts. Based on the model described above for a straight flight path instrument, the error on the evaluation of the field-of-view would be in the range of 10-12 % near the edges of the field-of-view, which translate into larger errors in the depth increment.

### 4.2 *Revisiting the linear projection*

Finite-element simulations of idealized smooth specimens have shown that the variability in the image compression factor across the field-of-view may be significant [20, 31]. Generally, atomistic simulations provide information on near field phenomena; however, boundary-element simulations were recently employed to investigate the impact of the specimen shape on the image compression factor [11]. The evolution of the image compression as a function of the ion launch angle, as shown in Figure 2 (a) for a specimen with a shank angle of two degrees, a radius of curvature of 50 nm, and for varying contact parameter, C, which is defined to be the sine of the angle between the shank edge and the tangent to the cap at the point of contact [11]. The contact parameter enables adjustment of the angle between the tangent to the spherical cap and the conical shank, which is zero when tangential continuity is satisfied and one in the case of a perfectly flat-top truncated cone (a more intuitive, but less flexible, parameter has been introduced previously by Larson et al. [59]).

Experimental values of contact parameter range from 0.18 to 0.49 (mean of 0.37) across more than 10 field evaporated specimens investigated from various materials incl. Si, Fe, Al and W. The variation in $\xi$ across the field-of-view may seem relatively small (<5% in this case) but at the specimen surface this small difference in image compression translates into an error on the order of several atomic spacings, which increases towards the edges of the field-of-view. Note, however, that the magnitude of this error may be limited when compared to the errors arising from near-field effects and to the uncertainty in the various other parameters (e.g. evaporation field, field factor, detection efficiency). We provide these results here to show that a point-projection cannot accurately reproduce the results of a linear projection over a wide field of view. Parameters extracted from correlative microscopy on atom probe specimens to deduce the radius of curvature or the shank angle do not necessarily yield the best reconstruction possible, although recent improvements have been made [59].

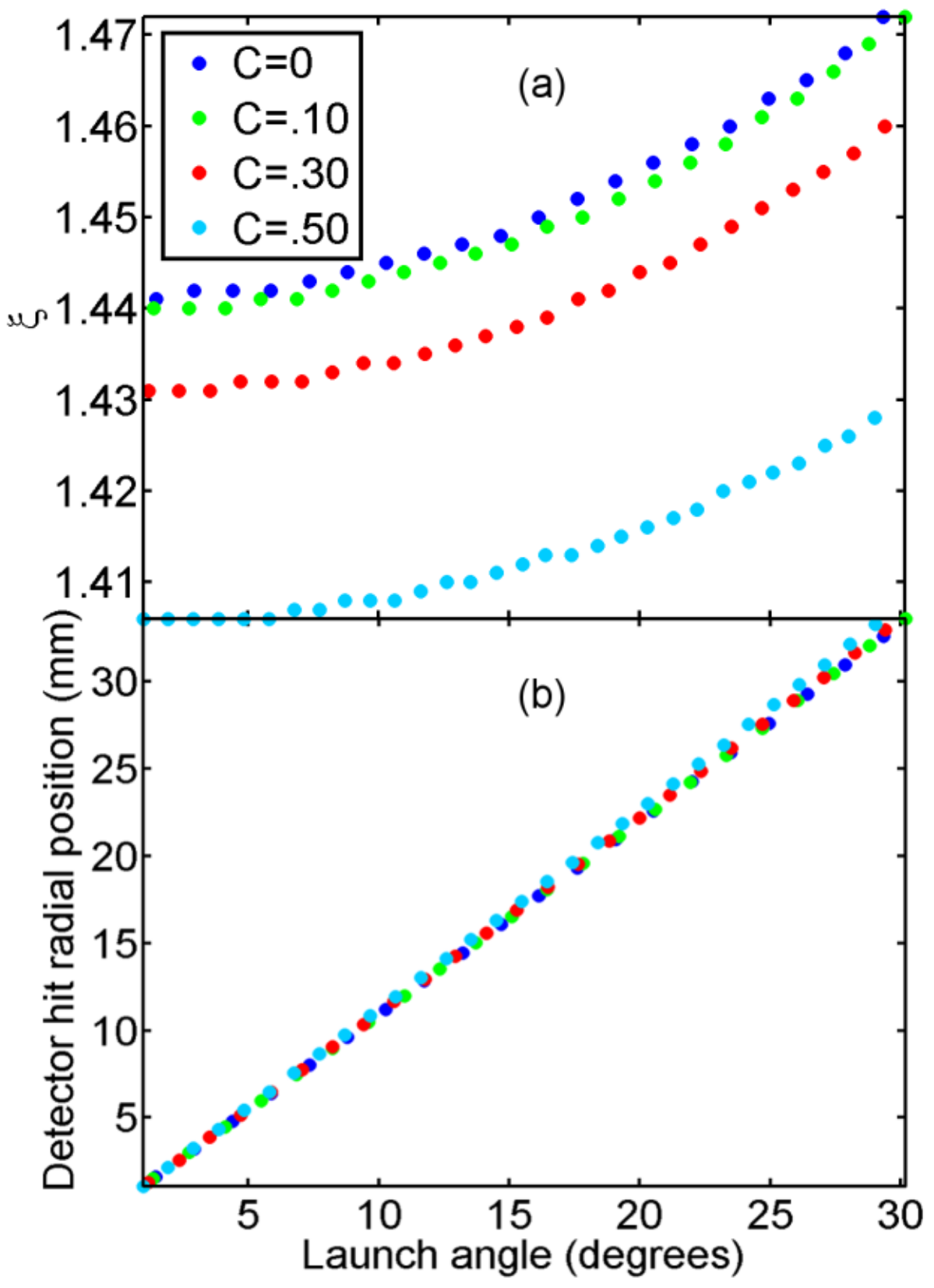

*Figure 2 (a) Image compression as a function of the launch angle for a specimen with a shank angle of 2 degrees, a radius of curvature of 50 nm, and for increasing contact parameter; (b) Plot of the detector radial position as a function of the launch angle for the same set of simulations.*

Ion hit position on the detector as a function of the launch angle for the same set of simulations is shown in Figure 2 (b). The results are similar to plots obtained by field ion microscopy to investigate the linear projection [46], and a linear relationship can be fitted with correlation parameters $R^2$ consistently above 0.999. The variation in the $k_\theta$ parameter does not appear substantial as the contact parameter increases. Furthermore, as shown in Figure 3, and as discussed by Humphry-Baker & Marquis [77], this relationship also holds for the use of the reflectron lens often found in modern instruments [78, 79]. From these perspectives, a linear projection appears robust over a variety of specimen shapes and instruments. Ultimately, one must bear in mind that ion trajectories are substantially influenced by near-field effects, mostly linked to the atomistic structure of the surface (e.g., crystallographic poles, different evaporation field phases, etc.) that would not be corrected by a change in the fundamental projection type. It is only from a far field perspective that the use of a linear projection has potential advantages over the standard point projection model.

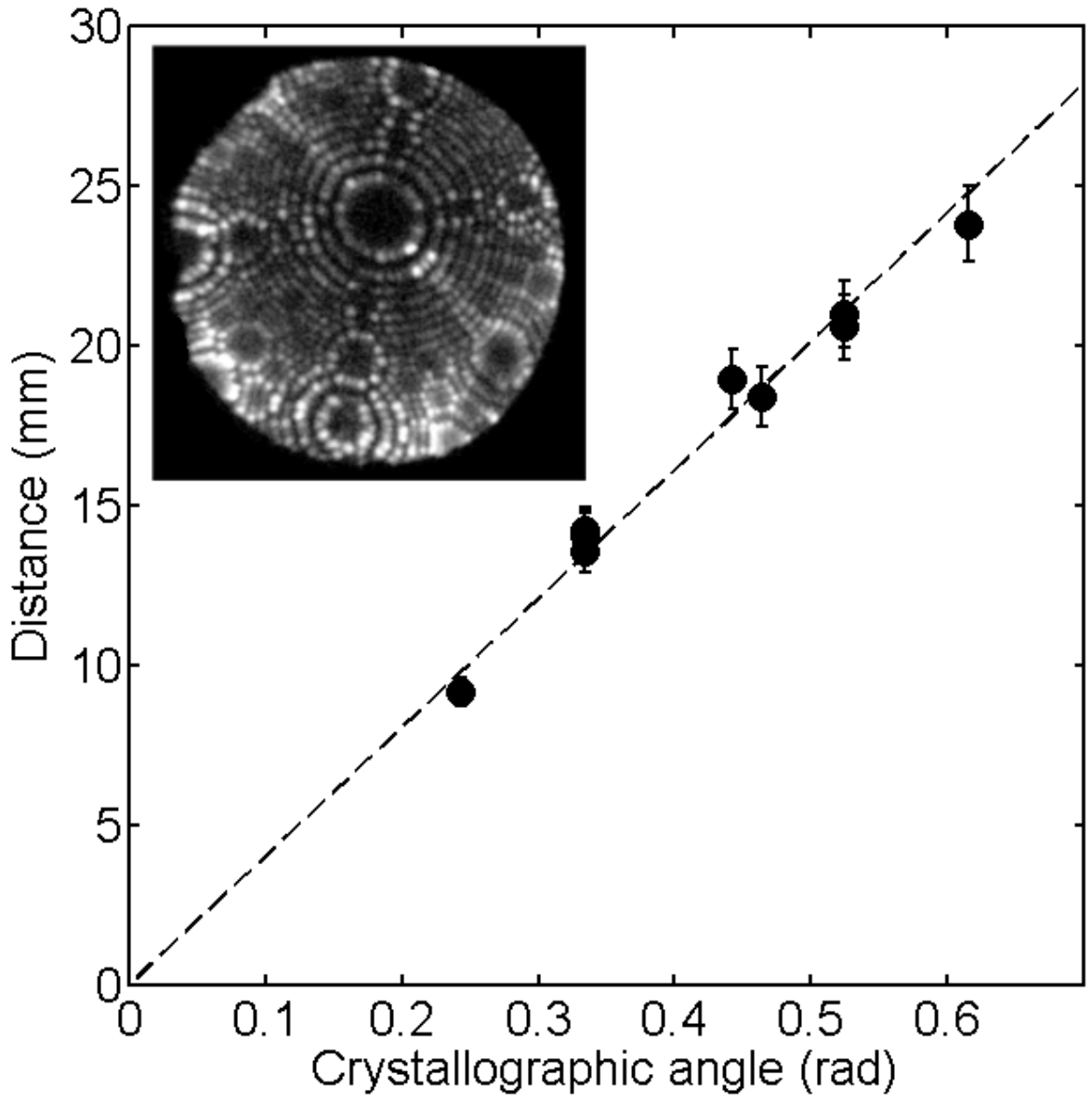

*Figure 3: Graph of the distance between features on the field ion micrograph as a function of the crystallographic angle between features. Inset shows the corresponding field ion micrograph of pure W obtained through the reflectron lens of a CAMECA LEAP 3000 HR.*

### 4.3 *Linear Projection with Dynamic Reconstruction*

Since the introduction of the standard model [24, 25], it has been assumed that the reconstruction parameters remain constant as the specimen evolves. However, there is known variability in the reconstruction parameters from specimen to specimen [43, 80]. The results of Loi et al. [11] reveal the dependence of such variables as $\xi$ and $k_f$, on specimen parameters such as radius and shank angle and it is hence reasonable to expect certain reconstruction parameters to vary during an experiment.

Gault et al. [81] recently proposed a method of reconstruction where the dataset is sectioned into subsets containing only 1–3 million ions. In each of these subsets, which represent a sequential sampling of the sequence of detection, an image compression factor is estimated based on the location of major crystallographic poles on the detector, and, subsequently, the subset is reconstructed and $k_f$ is optimized so as to obtain the appropriate interspacing for sets of atomic planes. The value of the two factors are recorded and an interpolation function is used to attribute, to every ion in the sequence, a unique combination of parameters then used to build a reconstruction of the complete dataset [81]. This dynamic reconstruction approach was shown to reduce the distortions compared to a protocol using all fixed parameters. This procedure, however, does have some drawbacks. First, its application is relatively labor intensive. Second, it is based on crystallographic information being obtained from the specimen (i.e., poles), which may not always be available.

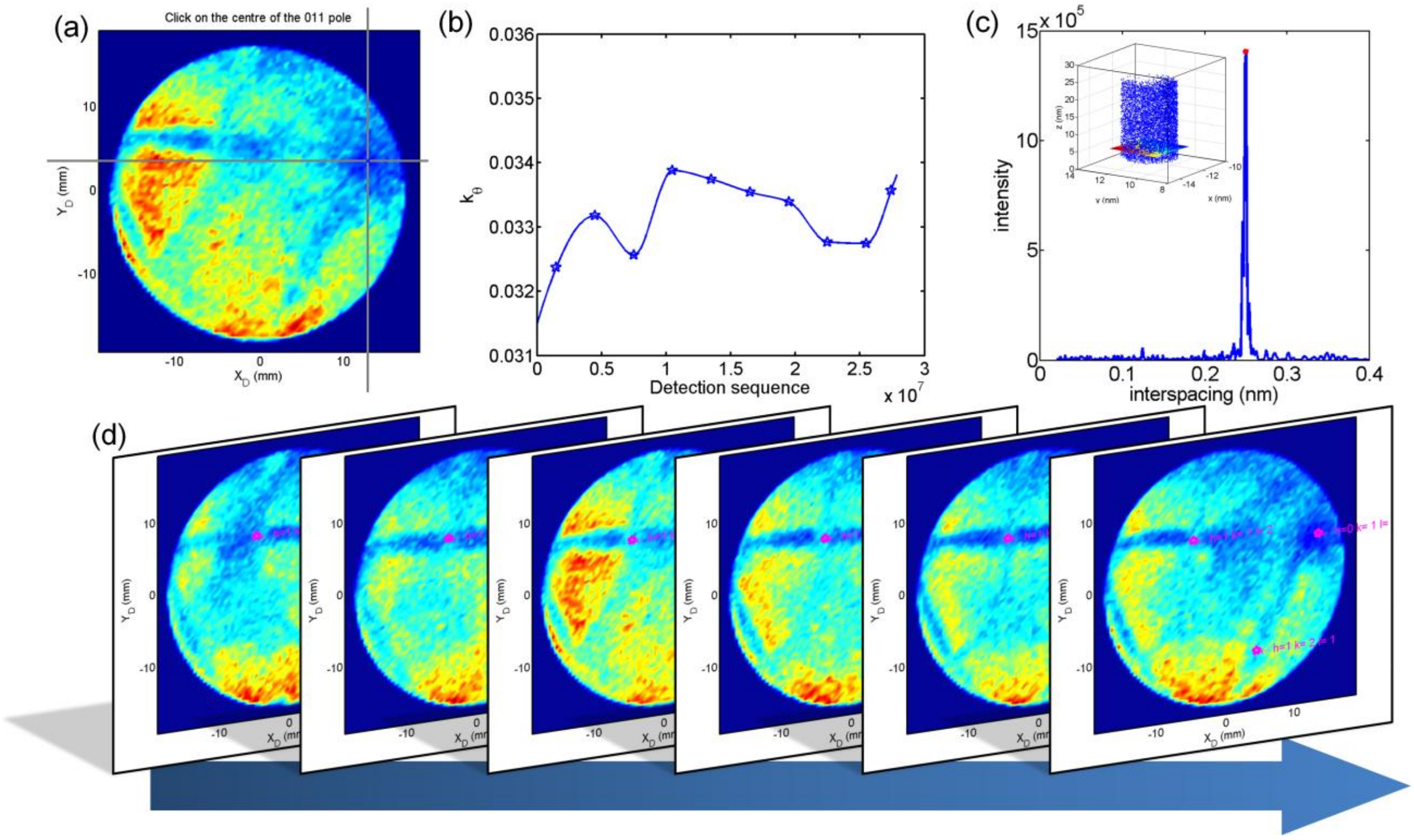

*Figure 4: Dynamic reconstruction protocol: (a) the coordinates of a selection of poles are interactively recorded by software from a desorption map within a segment of 3 million ions; (b) the* $k_\theta$ *is determined for this segment, based on the location of the poles and their crystallographic identity; (c) the interspacing of the planes around one or several poles within the reconstructed segment is determined by means of a FT, inset is shown the subset of the data and the fitted plane; (d) series of desorption maps through a dataset obtained on a ferritic steel illustrating the iterative and interactive protocol, the pink dots correspond to the location of the poles.*

An extension of the dynamic reconstruction protocol has recently been developed, which includes user input, as outlined in Figure 4. The dataset is sectioned and for each segment, a detector hit map histogram (sometimes referred to as desorption map) is created. On each map, the user indicates a set of crystallographic poles (Figure 4 (a)) which allow for a direct estimation of $k_\theta$ (Figure 4 (b)). The data section is then reconstructed based on this value and the value of the $k_f$ either defined manually (useful for the initial segment) or on the value of the preceding segment. The final step is the optimization of $k_f$ based on the measurement of the interspacing of atomic planes. In order to automate this step, a method based on a Fourier transform (FT) has been applied to extract the spacing of atomic planes from data centered on a crystallographic pole.

The steps in the FT-based method developed to measure the interspacing are: 1) a plane is fitted to the point cloud which, when atomic planes are imaged, provides the orientation of the planes within the subset, as shown inset in Figure 4 (c); 2) for each atom, the distance to this plane is calculated, and the values are stored in a matrix; 3) the FT is then applied to this matrix to reveal periodicity in the data and, as shown in Figure 4 (c) that displays the intensity of the FT as a function of the interspacing, a peak appears that allows to extract the value of the interspacing in the subset of the data.

This procedure is shown here to be efficient at picking planes and it may be applied to one or several poles. In the case where no planes can be visualized locally, due to the presence of a precipitate for instance, then the value of $k_f$ can be entered manually. Figure 4 (d) displays a series of desorption maps obtained during the reconstruction of a ferritic steel analysis by using this method. In order to further test its capability, this method was systematically applied using 3-nm diameter cylinders across a pure-Al dataset. Results are shown in Figure 5, in which the peak intensity in the FT reveals crystallographic poles within the dataset that can be correlated with poles from the detector hit map.

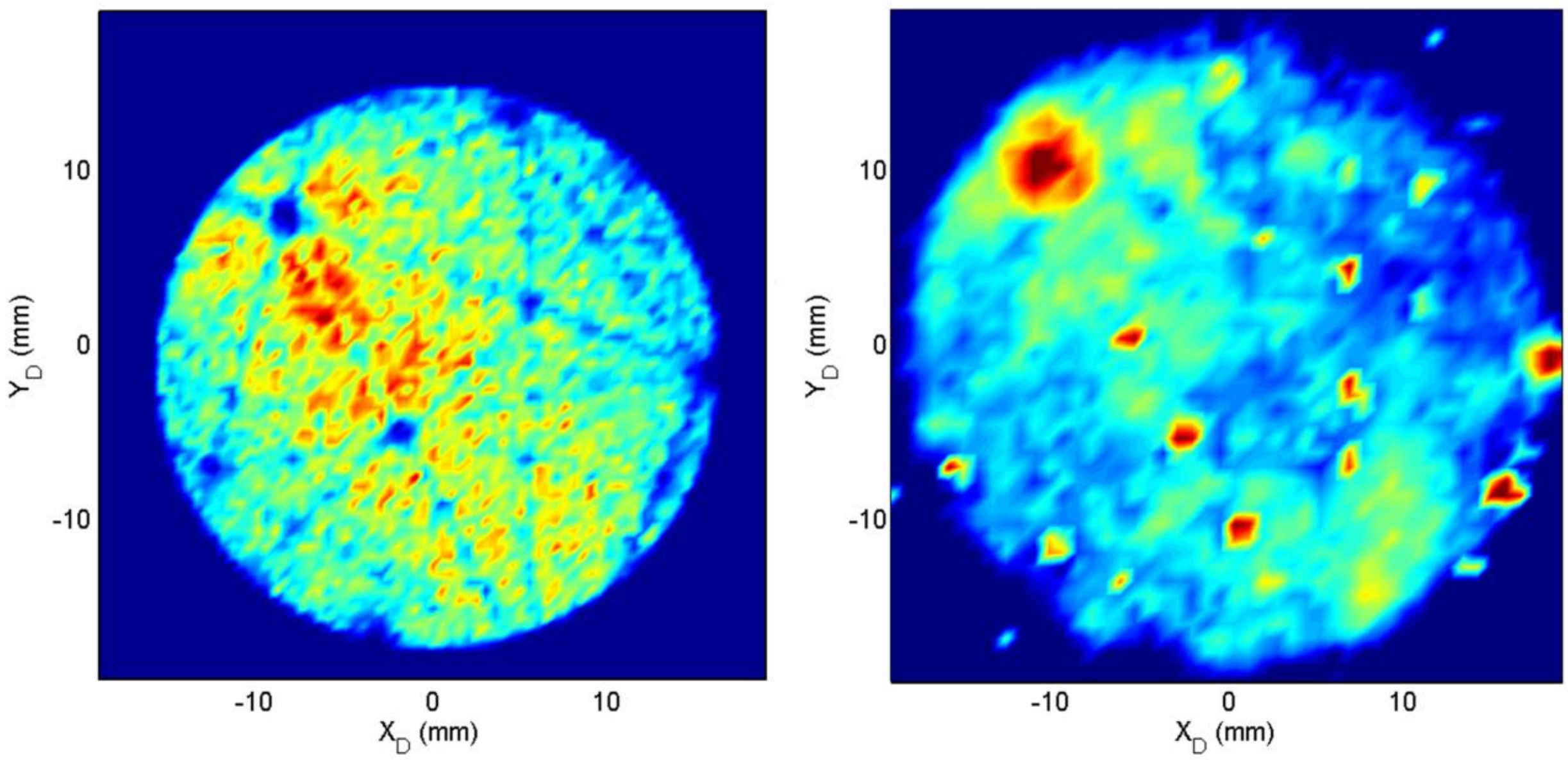

*Figure 5: Detector hit map (left) and corresponding intensity map (right) resulting from a systematic application of the FFT-based method to measure interspacing, revealing regions of high crystallinity within the dataset.*

The application of a dynamic reconstruction algorithm had previously demonstrated significant improvement [81]. An example is presented in Figure 6 for an AlMgSi alloy (analyzed with a reflectron-based CAMECA LEAP 3000X HR). The data in Figure 6 (a) were reconstructed using the standard (shank angle) method [27] after calibration of the depth using spatial distribution maps [82]. Calibration of the main projection parameter, $k_\theta$, by means of the interactive procedure described above, enables an adjustment of the parameters to produce the reconstruction displayed in Figure 6 (b). Although the differences between these datasets are subtle, the distortions on the edges of the dataset induce slight changes in the morphology of precipitates. This in turn modifies their potential for identification using conventional cluster identification methods that are based on inter-atomic distances. The data showed herein are an illustration of the promise for improved reconstruction based on the linear projection for general cases.

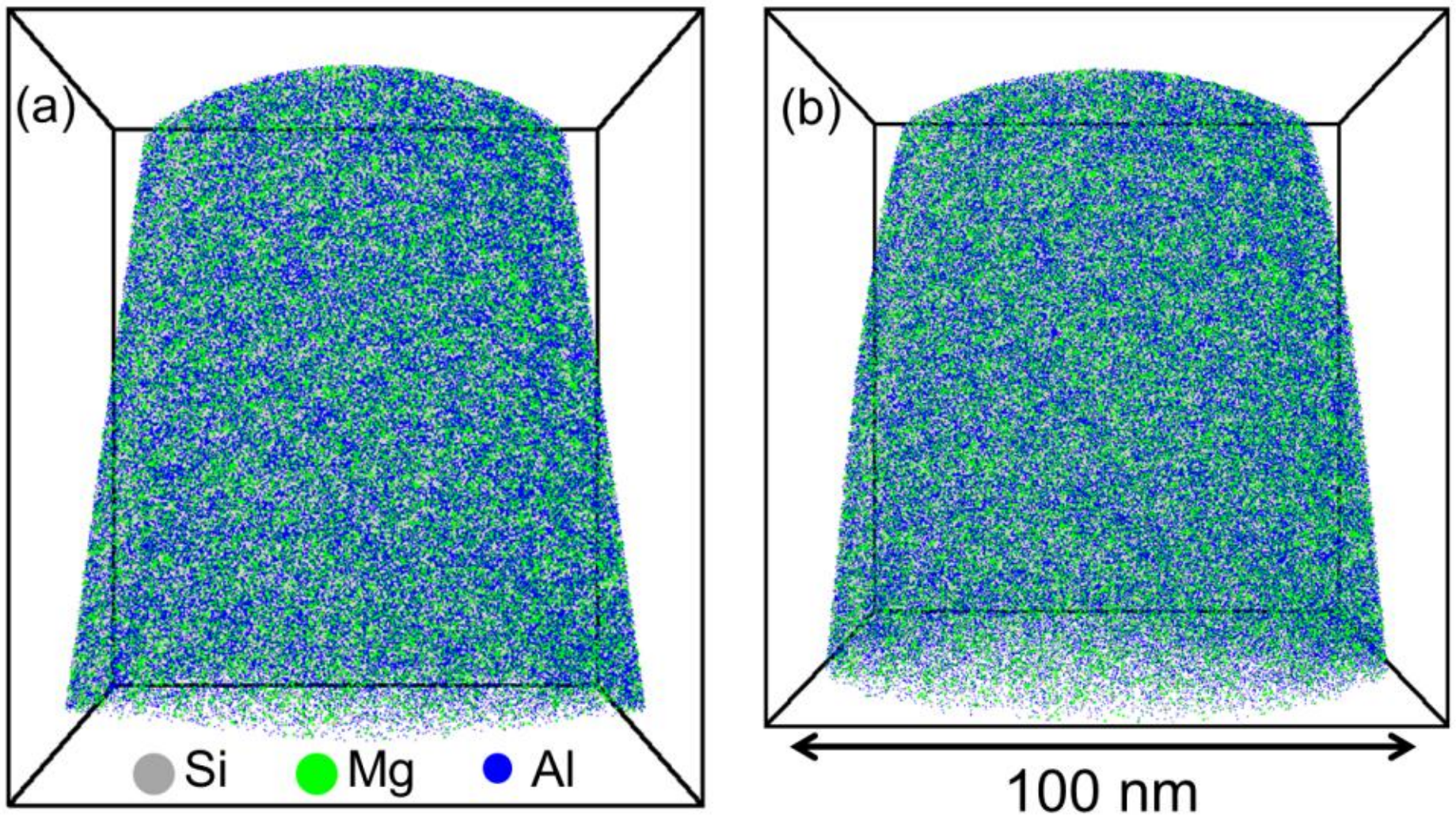

*Figure 6: (a) conventional reconstruction obtained using the conventional reconstruction algorithm from the analysis of an AlMgSi alloy, and (b) obtained via dynamic reconstruction.*

# 5 RESULTS & DISCUSSION: IN THE NEAR FIELD

## 5.1 *Introduction: Evolution Of The Specimen Shape*

In a homogeneous specimen, through successive field evaporation of atomic layers, the emitting surface at the apex tends to adopt a specific shape. As revealed by methods for measuring local radii, specimen shapes are non-spherical [83-85] (with atomic scale roughness in homogeneous materials due primarily to crystallography [62, 86, 87]). In the general case, the determining factors affecting this field evaporated shape are the crystallographic structure and orientation, the chemical composition of the specimen, and the specimen temperature. (Note that for the remaining discussion contained in the current work we will assume that temperature is low (~50K) and remains constant.) If the chemical composition is homogeneous (and the material is crystalline), then specimen reaches a steady-state shape comprised of a series of facets that depends on the local evaporation field (widely accepted as dependent on a combination of local bonding energy and work function at that location in the crystal [88-90]).

It is not, however, the specimen shapes induced by the presence of crystallinity in which we usually are most interested. From a materials science perspective we are often interested in microstructural features such as precipitates, interfaces, voids, grain boundaries, etc. and it is exactly these chemical or structural heterogeneities that most strongly affect the sequence in which atoms are field evaporated in atom probe tomography. These types of features locally modify bond strength and promote the development of local curvature [5], which impede the establishment of a smooth shape and complicate the reconstruction procedure.

Details of such a situation may be investigated by simulating a simple bilayer system with a moderate difference (20%) in evaporation field between the two layers [20, 22]. Plotting the ion hit position on the detector as a function of the ion sequence provides information regarding the uniformity of the field evaporation process, and thus an indication of whether simple or complicated shapes are likely to be forming. Note that in this type of figure at any point in time, the position of a layer of ions impacting the detector is a vertical line. For the bilayer case of a low-field material on a high-field material, Figure 7 (a), the detector hits are unevenly distributed, with significant local density variations arising as the evaporation proceeds from the upper layer (red) to the lower layer (blue). The lower layer requires a higher evaporation field and thus encounters a delay in its field evaporation upon being exposed at the surface. The blue is first exposed at the surface at the outer regions of the detector (white arrow) and final, when there is no low field region remaining, at the center of the detector (black arrow). Note that the coloring is simply attributed based on the sequence of detection, not according to the elemental nature of the field evaporated atom.

A comparison to experimental data is shown in Figure 7 (b) where field evaporation of pure aluminum shows nearly homogeneous density across the field-of-view in the ion sequence plot. Clearly the use of the same methodology to reconstruct the data for both cases will lead to artifacts at the interfaces in the bilayer example [22]. Quantifying the impact of local curvature on the magnification, the effects on the collected data, and its subsequent reconstruction, will be discussed in the following sections. The cases of precipitates and multi-layered systems will be used as examples.

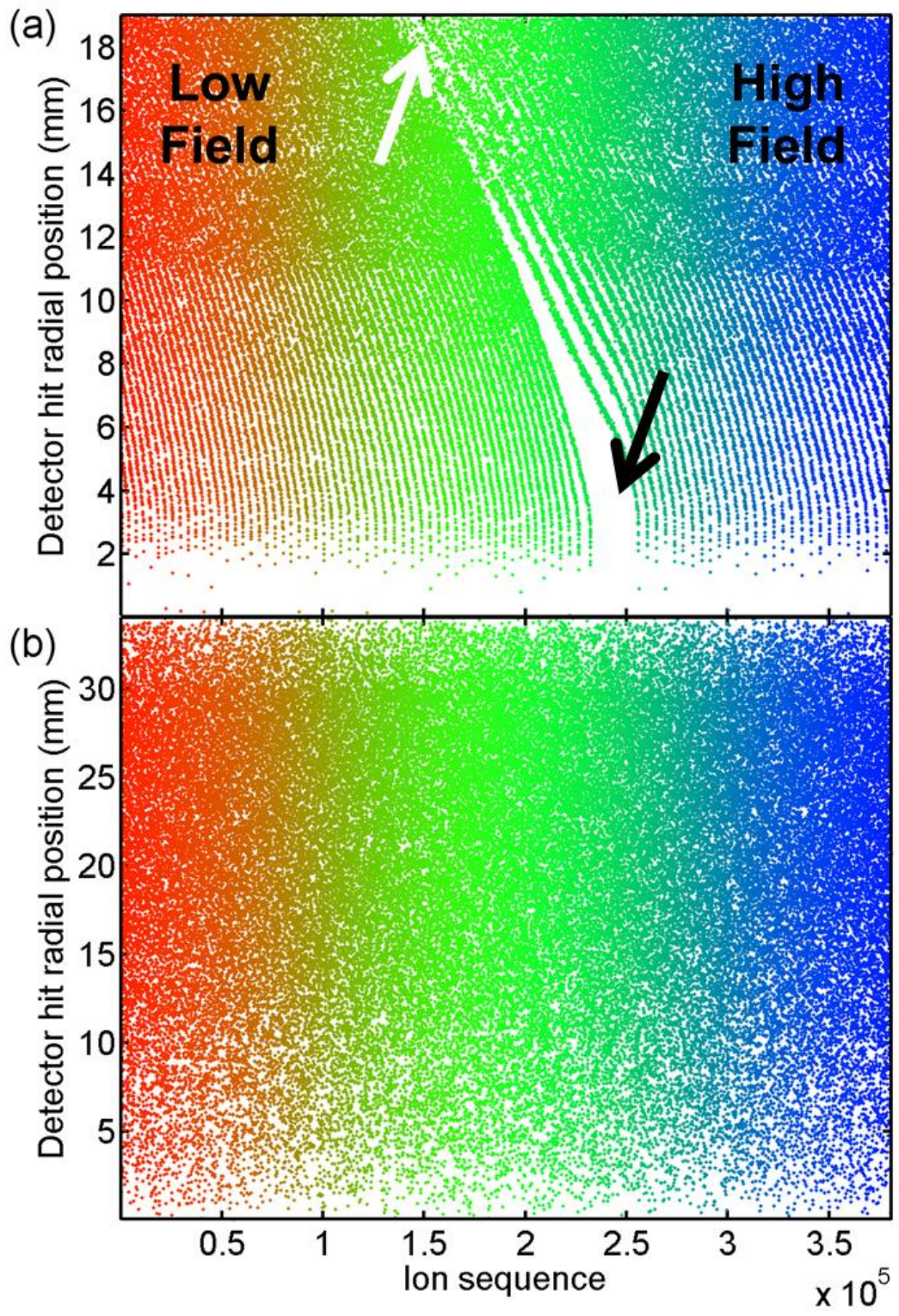

*Figure 7: (a) Detector hit radial position as a function of the ion sequence obtained during the simulation of an AB bi-layer where A has a lower evaporation field than B. Please note that the*

*low density in the range 0 – 2 mm on the detector is due to the presence of the (001) pole. (b) Similar plot for experimental data from the analysis of a pure Al specimen. Note that the vertical streaks clearly seen in (a), which are due to the field evaporation of individual atomic planes, cannot be seen in this dataset because the main crystallographic pole is off-centered.*

### 5.2 *Field Evaporation of Precipitates*

Solid-state precipitation of a given phase within a matrix is a key metallurgical phenomenon that is often used as a lever to optimize a material's properties and has hence been a topic of intense research using atom probe tomography (see [91-94] and the other papers in this volume for some recent reviews). Accurately reconstructing volumes containing precipitates can be challenging due to localized concentrations of atoms with a different evaporation field, which, as pointed out previously, lead to the development of local non-spherical curvature and the resulting local magnification and associated trajectory aberrations. Previous modeling work has been performed to investigate reconstruction artifacts associated with precipitates [7, 21, 65, 95], but our current understanding is still rather limited. In the following sections, we will discuss some efforts currently being pursued.

Using the full three-dimensional finite-element atomistic approach, we have simulated the cases of a spherical precipitate in a matrix, with the ratio of the evaporation field of the atoms within the precipitates to those within the matrix defined as defined as $\varepsilon$ (i.e., $E_P=\varepsilon E_M$). Two cases were investigated, first a low-field precipitate in a high field matrix with $\varepsilon$=0.85 and a high-field precipitate in a low-field matrix, with $\varepsilon$=2 (matrix evaporation field is half that of the precipitate evaporation field).

Consider the former case of a low evaporation field spherical precipitate as shown in Figure 8 (a). After some evaporation, as shown in the series displayed in Figure 8 (a–c), the specimen will form an apex structure whereby the precipitate region is flattened on the surface. In the series Figure 8 (d–f) depicting the opposite case (high-field precipitate), the situation is reversed, and once the second phase emerges at the surface, the entire endform is altered, with the high field phase clearly protruding from the surface. For both cases, once sufficient material is evaporated following the disappearance of the precipitate, the specimen resumes its equilibrium electrostatic endform.

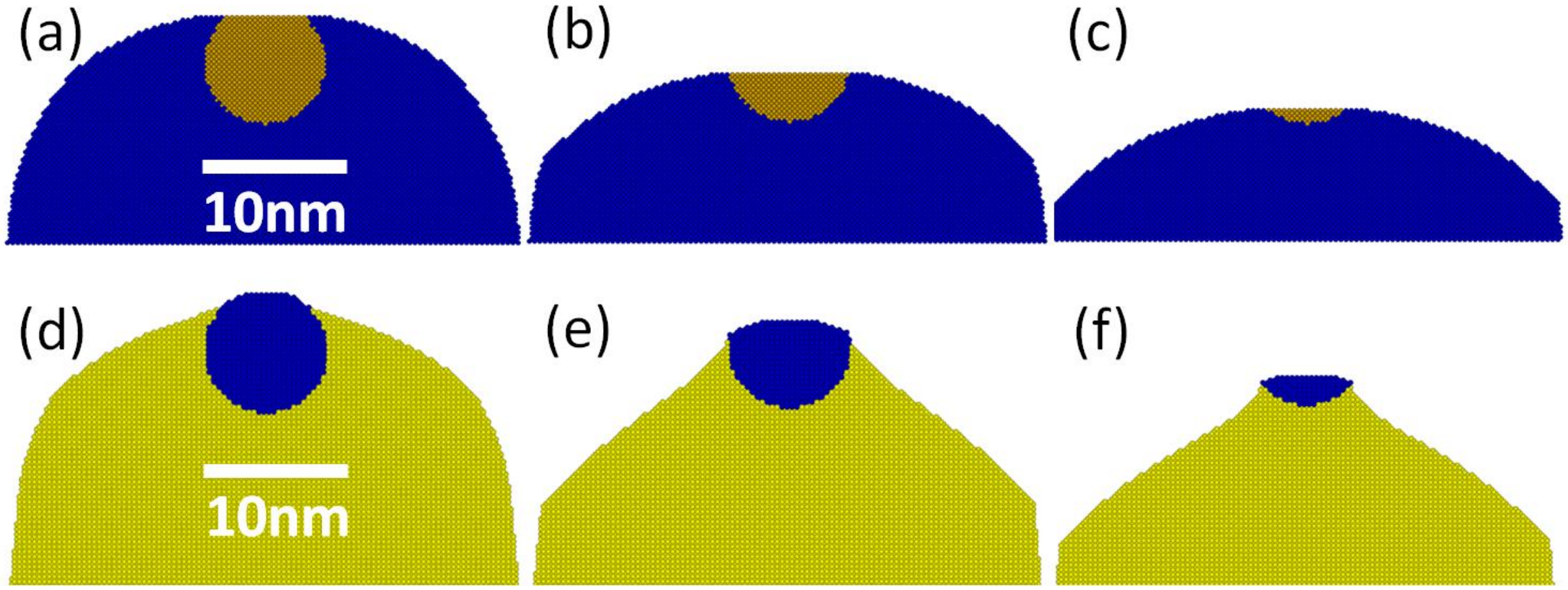

*Figure 8: Successive end form through the field evaporation of a specimen containing a single spherical precipitate. High evaporation field atoms are in blue, low evaporation field atoms in yellow. (a–c) sequence in the case of a low-field precipitate ($\varepsilon$=0.85); (d–e) similar sequence for a high-evaporation field precipitate ($\varepsilon$=2).*

The relative magnification between the precipitate and the matrix can be derived from plotting the impact positions as a function of the initial radial position of the atoms at the specimen surface, and the actual magnification for each phase may be estimated from the slope of the curve. These data displayed in Figure 9 (a) and (b).  are effectively a snapshot of the magnification during the field evaporation process of a single layer of atoms for a specimen containing the precipitates shapes presented in Figure 8 (b) and (e) respectively (reproduced inset). As expected, the trajectories of atoms in the matrix and the precipitate differ from the situation with no precipitates present in the simulations (solid black line).

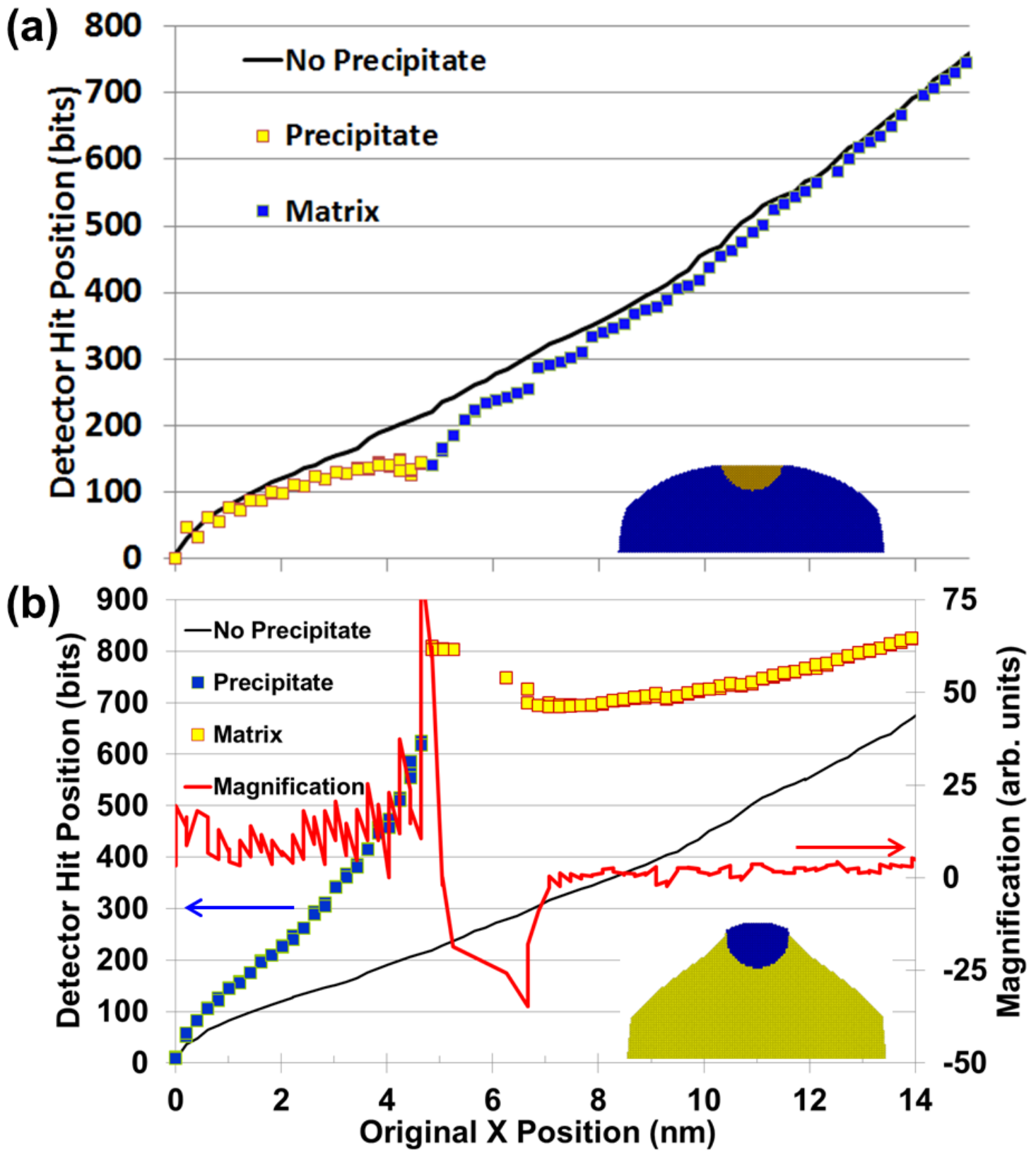

*Figure 9: detector hit position vs. original position at the specimen surface for the two shapes displayed on the top right of the image: (a) a low-field precipitate ($\varepsilon$=0.85) and (b) a high-field precipitate ($\varepsilon$=2). Note that x=0 represents the centre of the detector, and here also the centre of the specimen.*

The data in Figure 9 (a) suggest a low magnification (approaching zero slope) in the precipitate region (yellow points) with a narrow region of higher magnification in the matrix (blue points) near the precipitate. A small downward-pointing peak appears on the curve at the interface between atoms from the precipitate and the matrix. In this region the magnification changes very rapidly from low (in the precipitate) to high (in the matrix) and if this change is sufficiently rapid then atoms from both the matrix and the precipitate may overlap on the detector. This results in a region of artificial intermixing upon reconstruction [7]. The data in Figure 9 (b) exhibit the opposite trend, where the atoms of a high-field phase are subject to greater angular magnification than the low-field phase, as expected [5, 68, 95]. As shown in Figure 9 (b) the evolution of the magnification (red line) may be estimated by using the first derivative of the original atom position vs. detector high position data. The red line shows a very high magnification for the region of the precipitate near the edge of the matrix and a narrow region of negative magnification in the matrix very near the precipitate. This combination suggests the potential of ion crossing [96]. If ion paths do cross, the task of correcting this aberration gets more challenging and may be uncorrectable in a traditional ion-by-ion treatment.

The prominent role of the local curvature both at the microscopic level (i.e. atomic roughness, precipitates) and at a mesoscopic level (i.e. faceting, shank) on ion trajectories have long been known and discussed. However, a quantitative relationship is still missing and the true impact of local magnification and trajectory aberrations on the techniques performances is still difficult to assess [7, 21, 97].

The magnitude of the effects presented in this section are expected to vary significantly with two variables: 1) the evaporation field ratio $\varepsilon$, and 2) the ratio of the size of the precipitate relative to the specimen radius. The effects of these parameters are discussed in further detail in the next section from the perspective of furthering our understanding of local magnification near precipitates and the potential for ion crossing.

### 5.3 *Local Magnification Near Precipitates*

In order to investigate the phase space of the evaporation field ratio and the ratio of the size of the precipitate relative to the specimen radius, simulations have recently been performed using the 2D finite-element atomistic simulations described above. The 2D simulation assumes azimuthal rotational symmetry which aides in providing the computational speed required to more completely explore this phase space.

In the simulation, a cylindrical precipitate of radius $r_p$ is embedded within a matrix. This system adopts a nearly steady-state specimen shape although the non-zero shank angle does cause some increase in the radius during simulated field evaporation. Two examples $\varepsilon = 0.5$ and $\varepsilon = 2.0$ are shown in Figures 10 (a) and (b), respectively. In both cases, a schematic view of the cylindrical precipitate (yellow) in the core of the specimen has been superimposed so as to provide a clearer view of the actual simulated system. A dip forms at the apex due to the low evaporation field of the precipitate. In the latter, as expected, a protrusion develops at the specimen apex due to the high evaporation field of the precipitate. These results are consistent with Figure 8 and with previously reported literature results [65], confirming the accuracy of the 2D simulation.

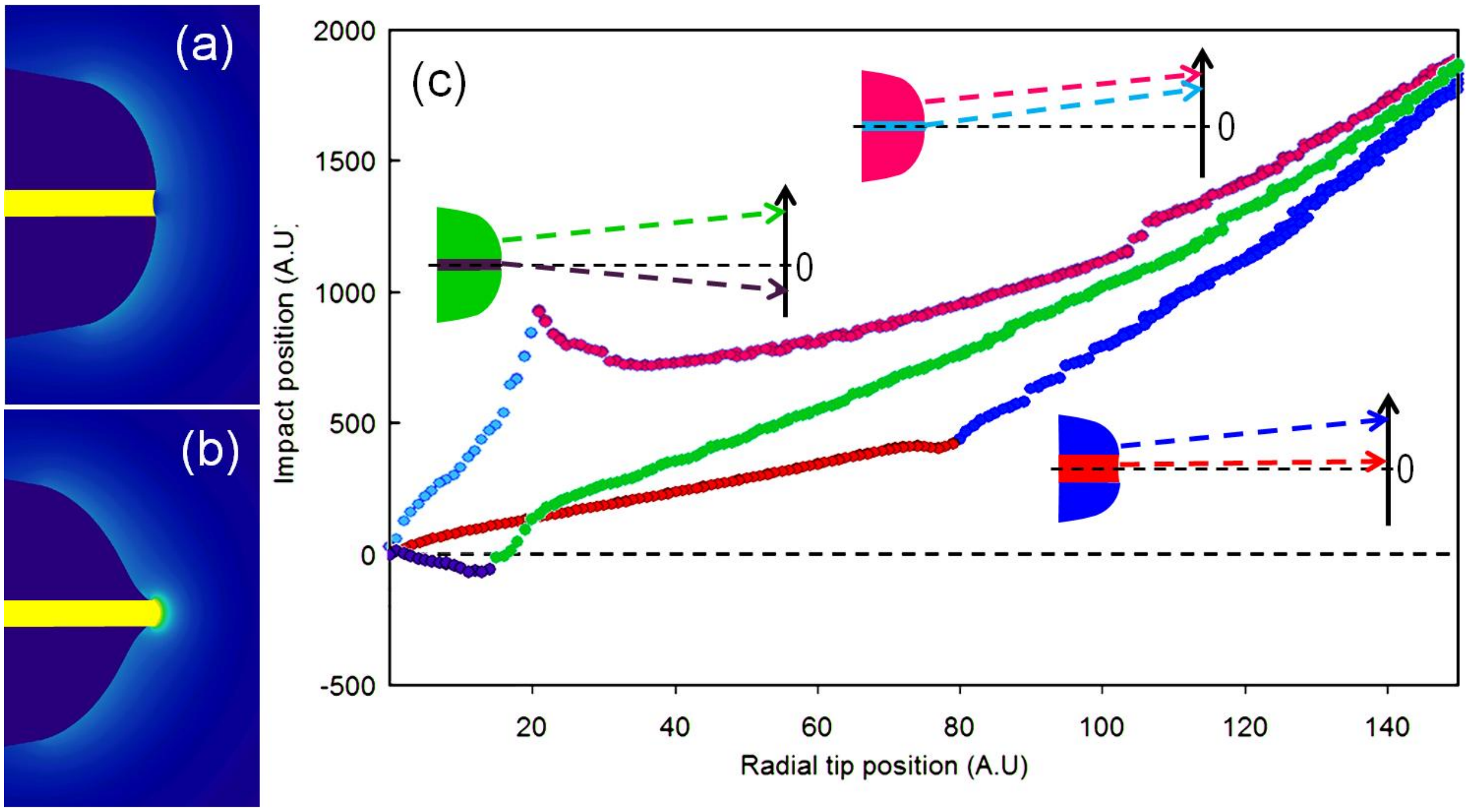

*Figure 10: (a) distribution of the electrostatic field around a simulated shape for ε=0.5 and, for clarity, a schematic view of the cylindrical precipitate in the core of the specimen has been superimposed (yellow); (b) similar distribution for ε=2.0 and, for clarity, a schematic view of the cylindrical precipitate in the core of the specimen has been superimposed (yellow); (c) Impact position on the detector as a function of the radial position at the specimen surface for cylindrical precipitate with $r_p$ ~3 nm and ε=2.0 (cyan and pink) , $r_p$ ~16nm and ε=0.55 (red and blue), and $r_p$ ~3 nm and ε=0.55 (purple and green). Schematic views shown in the insets.*

In Figure 10 (c), the 2D simulation data similar to Figure 9 are displayed, with detector hit impact position plotted as a function of the original atom position. It is readily apparent in the graph that the atoms from the precipitate (in cyan, purple, and red,) are subject to a different magnification than the atoms from the matrix (in magenta, green, and blue).

Let us first consider the case of the larger $r_p$ (~16nm) cylinder with ε=0.55 in red & blue. The precipitate has a lower evaporation field than the matrix, which translates into a slow-rising slope in comparison to the matrix and hence a compression of the projection of the precipitate. This compression induces a small local variation in the hit density on the detector and, more importantly, results in artificially high density regions within the standard model tomographic reconstruction. Let us now consider the second case, with a smaller precipitate radius $r_p$ ~ 3nm and a high evaporation field ε=2.0 (cyan and pink). Here the effect is opposite, a fast-rising slope for the atoms of the precipitate (cyan). This high local magnification region translates into a low atomic density in the standard model reconstruction. Then a negative slope is observed for the atoms of the matrix after the curve forms a peak (in pink) in the vicinity of the interface before reaching a steady increase. As discussed above, this combination of high magnification/negative magnification indicates a potential overlapping region between atoms of the precipitate and the matrix. A similar effect is seen in the third case ($r_p$ ~ 3nm and a low evaporation field, ε=0.55), where atoms from the precipitate (purple) and those from the matrix (green) overlap within the precipitate. A critical observation in this case is that the negative slope results in the ions landing on the opposite side of the detector (negative values). The aberrations not only lead to a deformation of the reconstruction of the precipitates, but again give rise to the potential for trajectory overlap and here result in an inversion of the image. These simulations throw light onto the issue of the potential for ion-crossing [96], the existence of which has been debated for years.

The phase space of the variables ε, and $r_p/R$ was explored in order to determine when ion crossing might be expected. In Figure 11, the relative magnification of ions in the precipitate, with respect those from the matrix, is plotted as a function of the size of the precipitate relative to the specimen radius of curvature. A negative value of the relative magnification indicates ion crossing, and hence image inversion. The results show that the latter appears below a critical value of the relative field evaporation of the precipitate ε (dashed line), which is also size dependent. Note that the simulation predicts the possibility of null magnification, which would cause a very high density region on the ion detector.

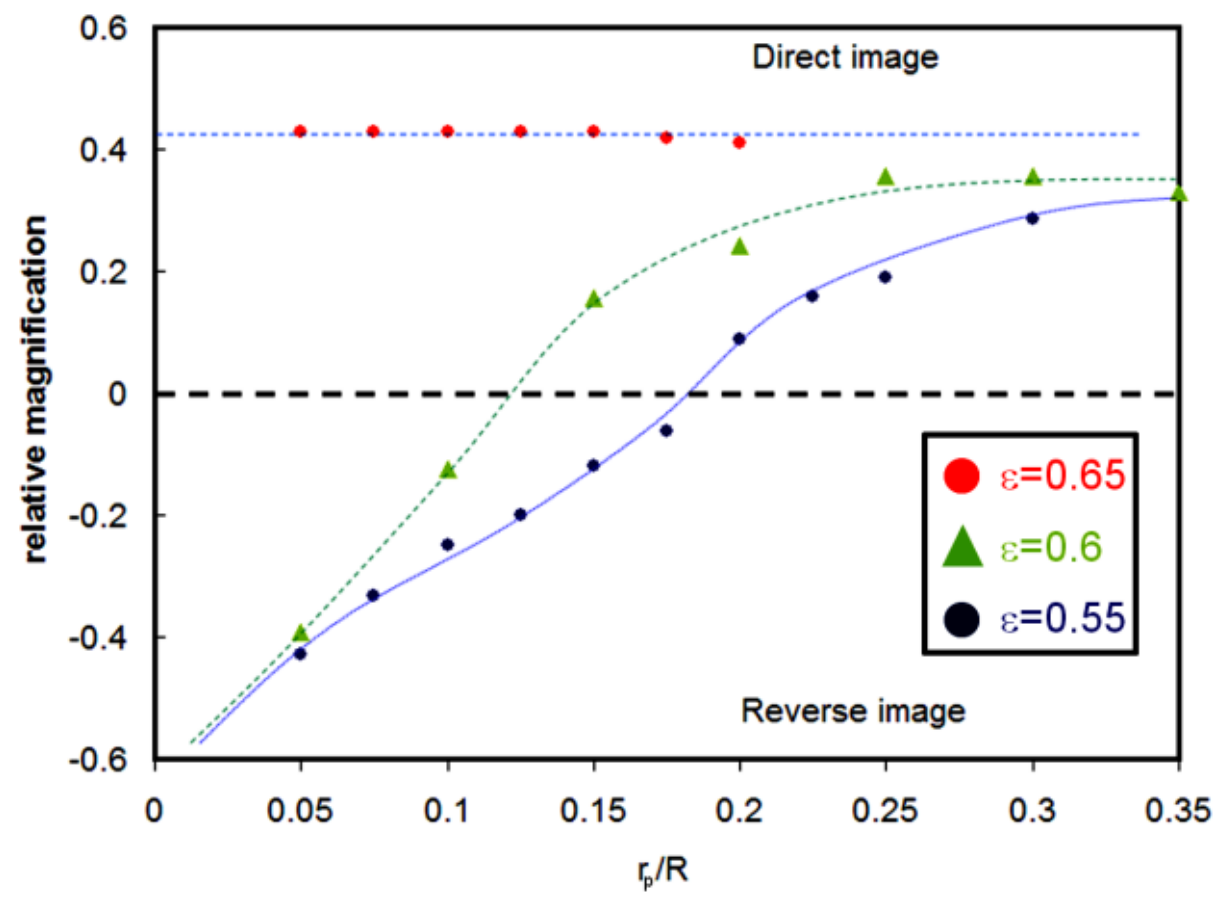

*Figure 11: Magnification measurements as a function of the size of the precipitate relative to the specimen radius ($r_p/R$) for three different value of ε. Inversion of the image is produced below a threshold of precipitate size, with ε sufficiently low.*

## 5.4 *Field Evaporation Of Multilayers*

In recent years, simulations have been extensively used to determine how the difference in evaporation field between phases or layers affects the establishment of the self-similar shape [3, 20-22, 75, 98]. Using plots similar to Figure 7 but with a color-code corresponding to the different atom types, it is possible to investigate the influence of the difference in the elemental evaporation field on the sequence of detection, as shown in Figure 12. The evaporation fields of the different layers ranks as such $F_B>F_D>F_C>F_A$. So from the perspective of the various interfaces encountered, the important considerations are $F_B>F_A$, $F_B>F_C$, and $F_D>F_A$.

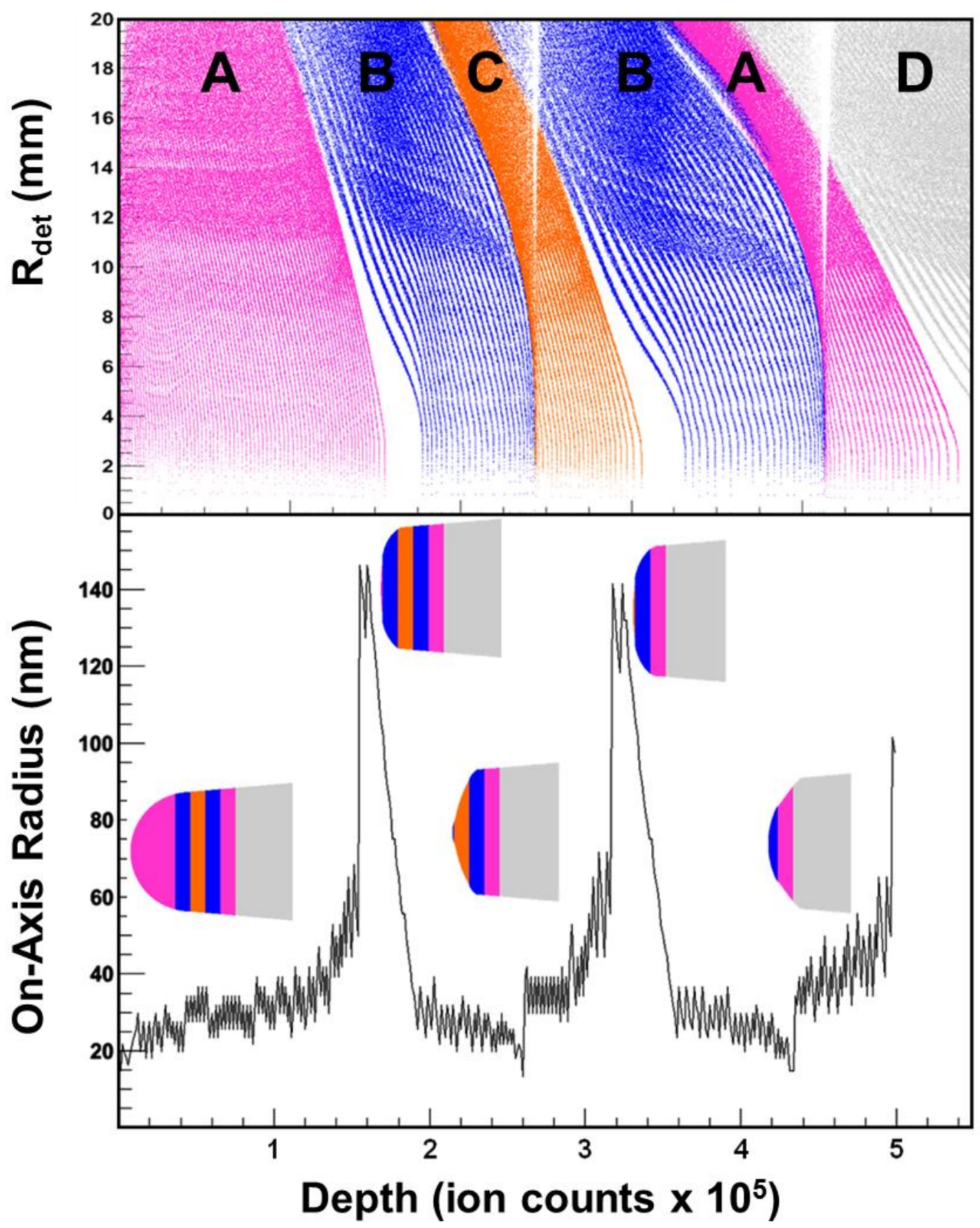

*Figure 12: (top) Detector hit radial position as a function of the ion sequence obtained during the simulation of an A-B-C-B-A-D multilayer system, and (bottom) the estimated on-axis radius measured over the same ion sequence.*

The radius of curvature at the apex of the specimen was measured through the evaporation of the different layers and is also included in Figure 12. From such an analysis performed on simulated data, we may immediately draw several conclusions:

- layers are not evaporated successively, but atoms are field evaporated depending on the local electric field and thus local curvature of the specimen
- local curvatures result in density variations on the detector

- a transition from a low to a high evaporation field layer (A to B, C to B, and A to D) results in a peak in the radius of curvature estimate
- a transition from a high to a low evaporation field layer (B to C and B to A) results in a drop in the radius curvature estimate

While the first two of these may appear somewhat obvious, the latter two are more interesting, and are counterintuitive: the low radius of curvature estimate arises for the low evaporation field species and the high radius of curvature arises for the high evaporation field species *in the interfacial regions*. This is opposite of the cases described above for precipitates. This illustrates the complexity of relating an actual measure of the specimen radius of curvature to the change in density.

In order to better understand the details of these effects, a method has been developed to trace the aberrations that occur upon reconstructing simulated data [31]. By positioning tracer planes within the original virtual specimen, it is possible to investigate the distortions caused by the shaping of the specimen during field evaporation coupled with the use of any model of reconstruction. The tracer planes of atoms should be reconstructed as either vertical or horizontal in a perfect reconstruction.

The case for bilayers with different evaporation fields is depicted in Figure 13 where the tracer planes are shown in white. In Figures 13 (a) and (b) a low-on-high interface in shown and for and (c) and (d) a high-on-low interface is shown (20% field difference). The deviations in the radial direction are shown in Figures 13 (a) and (c) while the z deviation is shown in Figures 13 (b) and (d). This type of simulated information is very useful in order to understand whether we should expect substantial aberrations in reconstructions of real data. In addition, the information may ultimately be used to improve the reconstructions – the tracer planes illustrate the errors in whatever reconstruction method is being employed, and this is of course what we are try to eliminate. Quantification of the distortions into three-dimensional functions may allow reconstructions to be corrected, in ways not dissimilar to other approaches already developed [6, 99, 100].

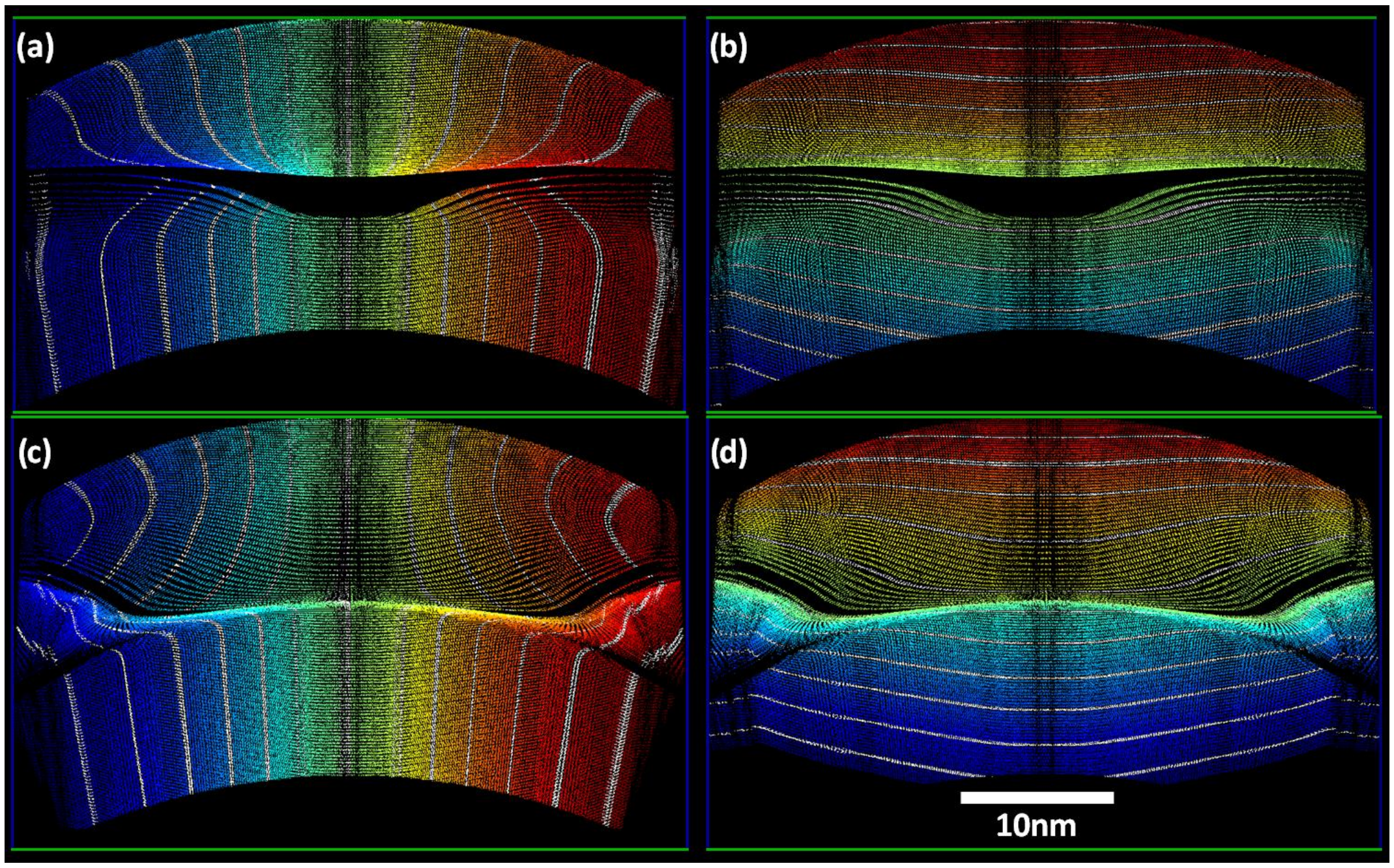

*Figure 13: Slice through a tomographic reconstruction obtained after the simulated field-evaporation of a bi-layer: (a) and (b) are for low-on-high interface and (c) and (d) are high-on-low (20% field difference). (a) and (c) are for the lateral (x,y) deviation and (b) and (d) are for variation in the analysis depth. Tracer planes, positioned vertically or horizontally and shown in white, were inserted to highlight the distortions both laterally and in depth, induced by the difference in the evaporation field between the two layers.*

### 5.5 *Simulation-based vs. simulation-informed reconstructions*

The first implementations of atomistic simulations have been a crucial reality-check that the effects induced by the combination of atomic roughness and local curvature across the specimen surface are extremely complex. A natural goal would be to re-compute the exact trajectory of each ion based on an atomically-defined specimen inserted within a simulation model, and hence allow overcoming the distortions induced by the simplicity of the projection used in the current paradigm. Haley et al. used a finite-element-based electrostatic computation of the distribution of the electric field generated by specimens [101], the shape of which had been extracted, in three-dimensions, using electron tomography techniques [102, 103]. They discussed the complexity of implementing such methods, which do not account for the detailed atomic structure of the emitting surface (near field). The computational cost of large-scale atomic-scale simulations, as well as the set of assumptions required to construct the simulation, preclude the implementation of such methods for the time being. However, simulations have demonstrated their merit beyond facilitating the interpretation of complex data [2, 21, 66, 98] and validating hypotheses with regards to fundamental aspects of field evaporation [2, 3, 7, 15, 19, 20, 22, 66, 97, 98, 104].

Simulations may have a unique role to play to inform the reconstruction protocol and, maybe more importantly, its evolution. Recently, Larson and Geiser et al. have introduced two simulation-informed paradigms that have potential to improve the reconstruction of bi-layer structures. In a first effort, they proposed to counteract the change in magnification caused by the varying shape the specimen through adjusting the image compression factor as a function of the radial position on the detector and varying this function with depth [20]. Other approaches, related to the work presented in sections 5.2 and 5.4, have demonstrated that the reconstruction of multilayer system may be improved by more accurately accounting for the evolution of the magnification through the field evaporation process [22]. Although these methods currently benefit from to the assumption of cylindrical symmetry in the simulated systems, there is no fundamental constraint that precludes their extension into three dimensions. However it needs to be pointed out that a direct translation of such information into a practical solution for improving the reconstruction of experimental data still remains extremely challenging.

## 6 Conclusions / Outstanding Challenges

We highlight here the possible avenues of development that currently will help bring the field of spatial data reconstruction in atom probe tomography to the next stage. These include 1) the use of simulations to inform the development of algorithms yielding more accurate reconstructions by better modeling experimental data, and 2) using simulation in the reconstruction algorithm itself by predicting the shaping of the specimen during field evaporation and quantifying and counteracting distortions. The methodology designed to reconstruct atom probe data needs to evolve to keep pace with the progress in other aspects of atom probe science and technology. A variety of field evaporation simulations illustrates a broad spectrum of effects, both near field and far field, which need to be corrected.

## Acknowledgements

The authors would like to thank our colleagues at CAMECA who assisted in assembling the materials presented in this manuscript and offered useful discussion, including T. J. Prosa, T. F. Kelly, J. D. Olson, D. A. Reinhard, P. H. Clifton, R. M. Ulfig, and E. Oltman. We extend our gratitude to Shyeh Tjing – Cleo – Loi who developed and performed the Lorentz-based simulations that have enabled part of this work. BG acknowledges that he is a full time employee of Elsevier Ltd. but declares no conflict of interest as his contribution to this article corresponds to activity during out of office hours. BG is grateful for the continuous support from Profs. S.P. Ringer & J.M. Cairney and the Australian Microscopy & Microanalysis Research Facility (AMMRF) at the University of Sydney. BG would like to thank Profs. M. P. Moody & C. R. M. Grovenor for their warm welcome at the Department of Materials at Oxford and the great honor of making me an academic visitor. L.T. Stephenson, R.K.W. Marceau, D. Haley, T.C. Peterson are all thanked for fruitful discussions over the years.

## FIGURE CAPTIONS

Figure 1: (a) Reverse-projection of a plane (dashed line) in a specimen with a radius of 50 nm, (initially projected using a linear law with $k_\theta$=0.056 rad m-1($\xi$=1.6, L=90 mm)), using the approximations of Bas et al. (red line), a true point projection with the input radius of curvature (dark blue line) and a point projection with a radius of curvature artificially increased to minimize the error (light blue line). Note the difference in scale between x and z. (b) Similar approach extended to a set of aligned spheres.

Figure 2 (a) Image compression as a function of the launch angle for a specimen with a shank angle of 2 degrees, a radius of curvature of 50 nm, and for increasing contact parameter; (b) Plot of the detector radial position as a function of the launch angle for the same set of simulations.

Figure 3: Graph of the distance between features on the field ion micrograph as a function of the crystallographic angle between features. Inset shows the corresponding field ion micrograph of pure W obtained through the reflectron lens of a CAMECA LEAP 3000 HR.

Figure 4. Dynamic reconstruction protocol: (a) the coordinates of a selection of poles are interactively recorded by software from a desorption map within a segment of 3 million ions; (b) the $k_\theta$ is determined for this segment, based on the location of the poles and their crystallographic identity; (c) the interspacing of the planes around one or several poles within the reconstructed segment is determined by mean of a FT, inset is shown the subset of the data and the fitted plane; (d) series of desorption maps through a dataset obtained on a ferritic steel illustrating the iterative and interactive protocol, the pink dots correspond to the location of the poles.

Figure 5. Detector hit map (left) and corresponding intensity map (right) resulting from a systematic application of the FFT-based method to measure interspacing, revealing regions of high crystallinity within the dataset.

Figure 6 (a) Conventional reconstruction obtained using the conventional reconstruction algorithm from the analysis of an AlMgSi alloy, and (b) obtained via dynamic reconstruction.

Figure 7 (a) Detector hit radial position as a function of the ion sequence obtained during the simulation of an AB bi-layer where A has a lower evaporation field than B. Please note that the low density in the range 0 – 2 mm on the detector is due to the presence of the (001) pole. (b) Similar plot for experimental data from the analysis of a pure Al specimen.

Figure 8: Successive end form through the field evaporation of a specimen containing a single spherical precipitate. High evaporation field atoms are in blue, low evaporation field atoms in yellow. (a–c) sequence in the case of a low-field precipitate ($\varepsilon$=0.85); (d–e) similar sequence for a high-evaporation field precipitate ($\varepsilon$=2.0).

Figure 9: detector hit position vs. original position at the specimen surface for the two shapes displayed on the top right of the image: (a) a low-field precipitate ($\varepsilon$=0.85) and (b) a high-field precipitate ($\varepsilon$=2.0).

Figure 10: (a) distribution of the electrostatic field around a simulated shape for $\varepsilon$=0.5 and, for clarity, a schematic view of the cylindrical precipitate in the core of the specimen has been superimposed (yellow); (b) similar distribution for $\varepsilon$=2.0 and, for clarity, a schematic view of the cylindrical precipitate in the core of the specimen has been superimposed (yellow); (c) Impact position on the detector as a function of the radial position at the specimen surface for cylindrical

precipitate with $r_p$ ~3 nm and ε=2.0 (cyan and pink) , $r_p$ ~16nm and ε=0.55 (red and blue), and $r_p$ ~3 nm and ε=0.55 (purple and green). Schematic views shown in the insets.

Figure 11: Magnification measurements as a function of the size of the precipitate relative to the specimen radius (rp/R) for three different value of ε. Inversion of the image is produced below a threshold of precipitate size, with ε sufficiently low.

Figure 12: (top) Detector hit radial position as a function of the ion sequence obtained during the simulation of an A-B-C-B-A-D multilayer system, the estimated on-axis radius measured over the same ion sequence.

Figure 13: Slice through a tomographic reconstruction obtained after the simulated field-evaporation of a bi-layer: (a) and (b) are for low-on-high interface and (c) and (d) are high-on-low (20% field difference). (a) and (c) are for the lateral (x,y) deviation and (b) and (d) are for variation in the analysis depth. Tracer planes, positioned vertically or horizontally and shown in white, were inserted to highlight the distortions both laterally and in depth, induced by the difference in the evaporation field between the two layers.